# Optical properties of $R$Cd$_3$P$_3$ ($R$: Ce or La) compounds: Insulator–metal transition induced by displacement of atoms in the unit cell

Jaekyung Jang[1,†], Yu-Seong Seo[1,†], Jeonghun Lee[2], Eundeok Mun[2], and Jungseek Hwang[1,3,*]

[1]*Department of Physics, Sungkyunkwan University, Suwon, 16419, Republic of Korea*

[2]*Department of Physics, Simon Fraser University, Burnaby, BC V5A 1S6, Canada*

[3]*Department of Physics and Astronomy, McMaster University, Hamilton, ON, L8S 4M1, Canada*

[†]*These authors contributed equally to this work.*

[*]*Corresponding author:* +82-31-299-4545 and jungseek@skku.edu

Abstract

We examined the electronic structures and optical properties of single crystals of $R$Cd$_3$P$_3$ ($R$ = Ce or La). Our first-principles analysis indicates that CeCd$_3$P$_3$ and LaCd$_3$P$_3$ exhibit semiconductor characteristics with narrow energy gaps of approximately 0.51 and 0.70 eV, respectively. Notably, a slight displacement of the Cd and P atoms within the unit cell significantly transforms the electronic structure from insulating to metallic state. Optical spectroscopy of both compounds reveals a metallic state with a low charge carrier density, suggesting a finite density of states at the Fermi level. A comparison between the theoretical electronic structures and experimental optical properties elucidates the observed metallic behavior. Additionally, the notable modification of the infrared-active phonons strongly indicates a structural phase transition in these compounds. Our findings also suggest that CeCd$_3$P$_3$ serves as a suitable platform for investigating the photoinduced Kondo effect due to its metallic ground state with limited charge carriers.

**INTRODUCTION**

Various quantum phenomena arising from geometrically frustrated spin systems continue to drive the investigation of quantum materials with two-dimensional triangular lattices likely to exhibit antiferromagnetic interactions[1-3]. A high degree of geometric frustration in such triangular lattices can lead to degenerate ground states and enhance the likelihood of observing unconventional quantum ground states. Magnetic frustration, combined with increased quantum fluctuations, can inhibit long-range magnetic ordering and result in magnetic liquid (spin liquid) states[4-12]. Among various materials, the spin-orbit coupling (SOC) in triangular lattice systems with $4f$ electrons intensifies quantum fluctuations and promotes the spin–liquid ground state. For example, the spin–liquid state has been proposed for insulating triangular magnets with $4f$ electrons, such as $NaYbSe_2$, $NaYbO_2$, and $YbMgGaO_4$[13-28].

In the past decade, numerous theoretical and experimental studies have focused on investigating $RT_3X_3$ ($R$ = rare earth, $T$ = Al, Zn, and Cd, and $X$ = C, P, and As) compounds with a hexagonal $ScAl_3C_3$-type structure ($P6_3/mmc$). In this structure, the trivalent rare-earth ions form a frustrated triangular lattice. Owing to the separation of the $R$ triangular layer by $T$ and $X$ atoms, it can be treated as a two-dimensional system[29-36]. Among these materials, the polycrystalline $CeCd_3P_3$ compound was suggested to exhibit quantum spin liquid behavior, as no magnetic ordering has been observed at temperatures as low as 0.48 K[31]. In contrast, physical property measurements of $CeCd_3P_3$ single crystals revealed antiferromagnetic ordering below 0.41 K[32]. This family of compounds has garnered significant interest due to its diverse electronic structures, which are significantly influenced by the sample preparation methods. Electrical and Hall resistivity measurements of single crystals of $LaCd_3P_3$ and $CeCd_3P_3$ grown via the flux method exhibited metallic behavior with a low carrier density, while polycrystalline samples exhibited semiconducting behavior with a small energy gap. Similarly, for $RCd_3As_3$ compounds, polycrystalline samples and single crystals grown through chemical vapor transport showed semiconducting or insulating behavior[34]. Conversely, single crystals of $RCd_3As_3$ grown using the flux method demonstrated metallic behavior[35]. In addition, it has been shown that the Kondo effect can be induced in a $p$-type semiconductor, $CeZn_3P_3$, by illuminating visible light[29,30]. In the case of $RZn_3P_3$ compounds, a trend from semiconducting to metallic states is observed as the rare-earth series progresses from La to Gd (decreasing $R$ radius), indicating a delicate balance between the unit cell volume and band structure near the Fermi level. Notably, within the $RT_3X_3$ family, only metallic samples exhibit a (potentially structural) phase transition at elevated temperatures, with detailed electronic structure calculations yet to be reported.

In this study, we examined the electronic structures and optical properties of $R$Cd$_3$P$_3$ ($R$ = Ce and La) compounds. Theoretical calculations revealed that CeCd$_3$P$_3$ and LaCd$_3$P$_3$ are semiconductors with energy gaps of approximately 0.51 and 0.70 eV, respectively. However, the measured optical spectra of CeCd$_3$P$_3$ and LaCd$_3$P$_3$ exhibited metallic behavior with a low charge carrier density, attributed to the finite density of states at the Fermi level. Theoretical calculations indicated that adjusting the Wyckoff positions of the Cd1 and P1 atoms in the crystal structure could lead to an insulator-to-metal transition in CeCd$_3$P$_3$ and LaCd$_3$P$_3$, highlighting the sensitivity of the electronic structure to subtle atomic displacements in tetrahedrally coordinated Cd1 and trigonally coordinated Cd2 environments. The observed metallic behavior in experiments may be attributed to subtle changes in the Cd1 and P1 atoms within the crystal structure. Notably, a significant alteration in the phonon modes in the infrared (IR) region suggested a structural phase transition in these compounds, with a phonon mode splitting in the optical spectra observed below the anomaly temperature ($T_s$) identified in the resistivity and specific heat.

**RESULTS AND DISCUSSION**

Figures 1 (a) and (b) show the measured reflectance spectra of CeCd$_3$P$_3$ and LaCd$_3$P$_3$ at different temperatures. Both samples display similar reflectance levels and shapes throughout the measurement range, suggesting comparable electronic structures and analogous phonon modes in the far-IR region. With decreasing temperature, both samples exhibit an increase in reflectance spectra, particularly below ~400 cm$^{-1}$ in the far-IR region, which is a characteristic optical trait of typical metals. However, the sharp rise in reflectance for these compounds is significantly lower than that of standard metals. For example, aluminum exhibits a sharp reflectance rise in the ultraviolet region, known as the plasma reflection edge, which is closely linked to the charge carrier density.

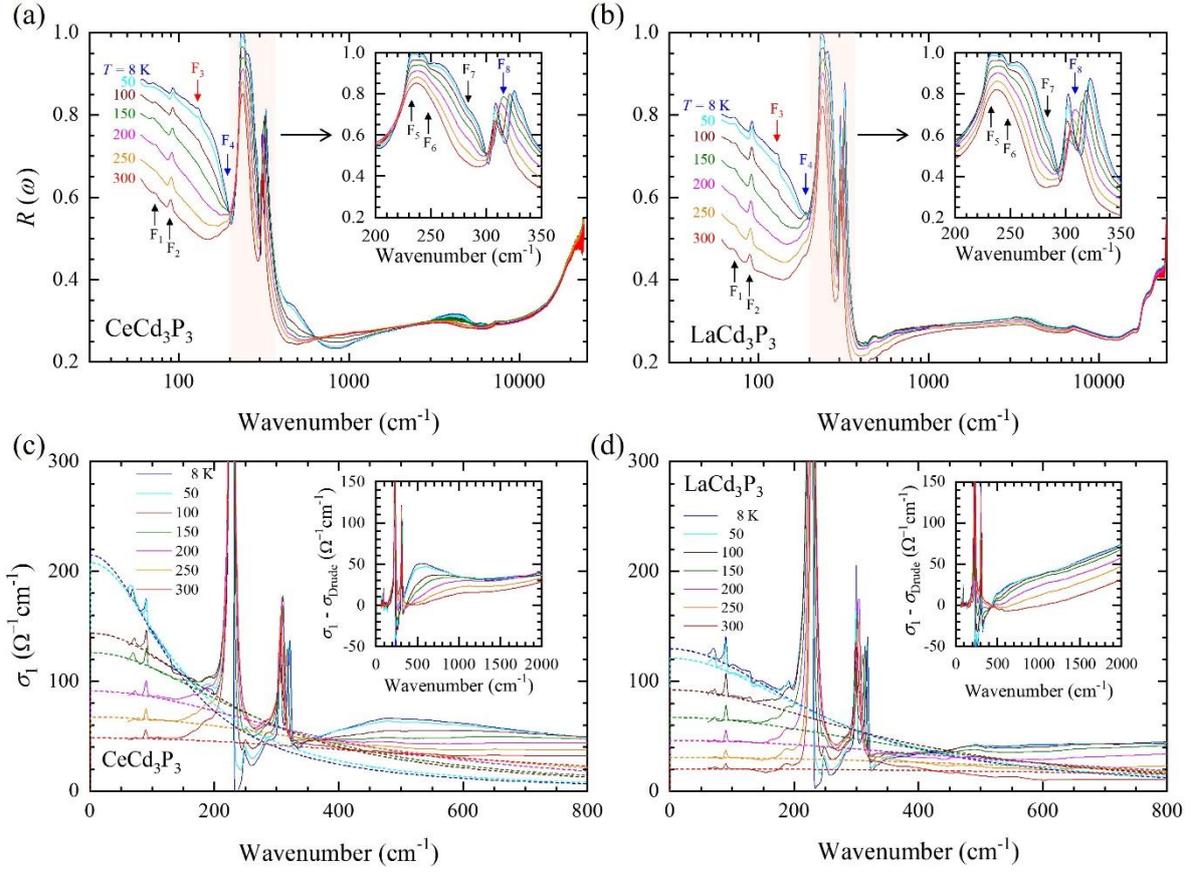

**Figure 1**. Measured *ab*-plane reflectance spectra of (a) $CeCd_3P_3$ and (b) $LaCd_3P_3$ at different temperatures. The insets provide an expanded view of phonon modes in the 200–350 cm$^{-1}$ range. Optical conductivity of (c) $CeCd_3P_3$ and (d) $LaCd_3P_3$ at corresponding temperatures, with dashed lines representing the Drude model fits (see details in the text). The insets display the optical conductivity after subtracting the Drude component.

The real parts of the optical conductivity spectra of both samples were derived from the measured reflectance spectra using the Kramers–Kronig analysis[37,38] and are depicted in Figs. 1 (c) and (d). Distinct IR-active phonon modes are observed within the 200–350 cm$^{-1}$ range. The optical conductivity spectra of both compounds exhibited finite values near zero wavenumber. Below 200 cm$^{-1}$, the optical conductivity magnitude increased consistently with decreasing temperature, suggesting metallic electronic ground states. However, the optical conductivity in the low-energy region was notably lower than that of typical metals, indicating a significantly smaller charge carrier density compared to standard metals. It is noteworthy that the optically estimated DC resistivities of $CeCd_3P_3$ and $LaCd_3P_3$ at 300 K are approximately 22 and 50 mΩ cm, respectively, which are three orders of magnitude higher than that of typical metals.

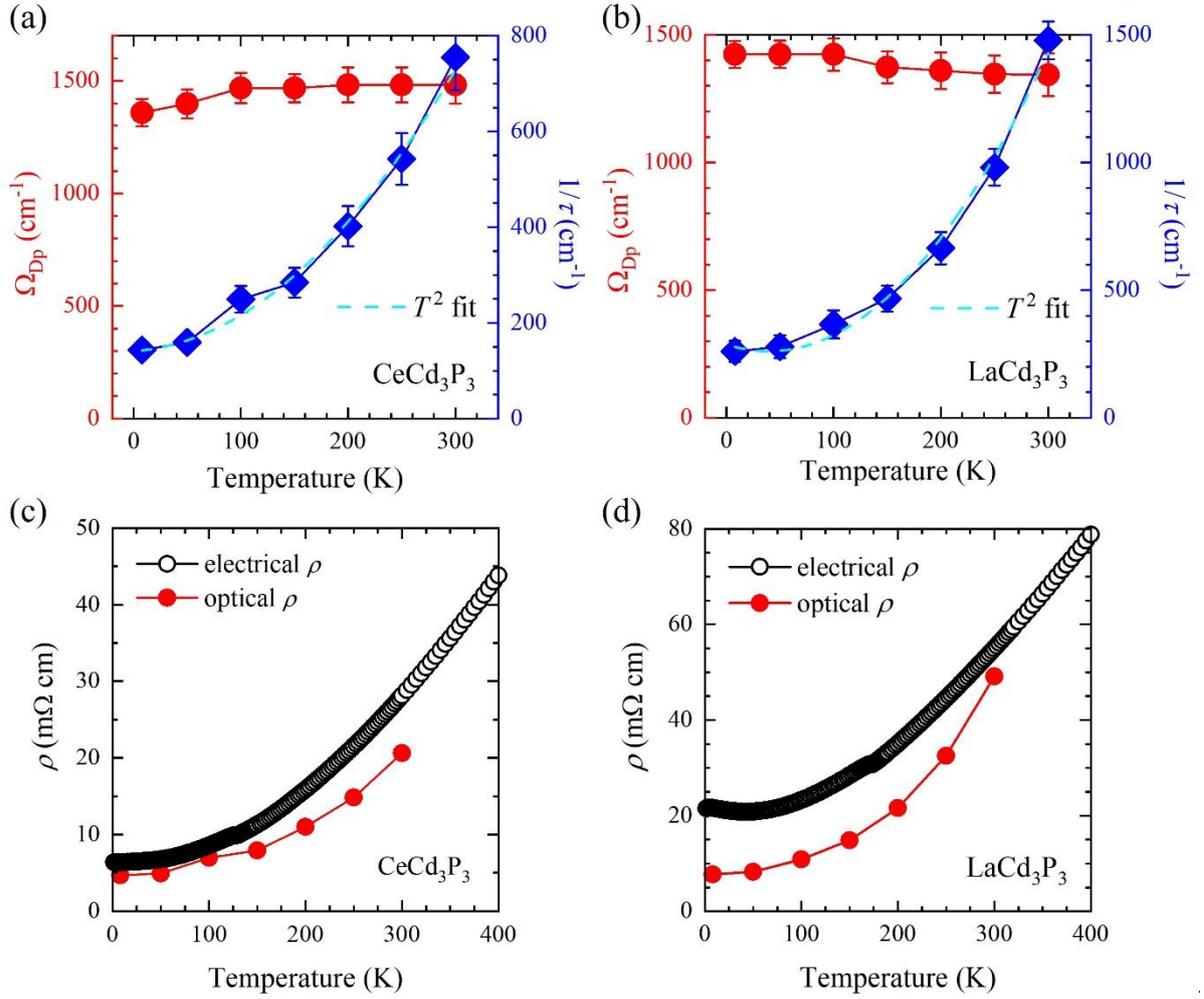

**Figure 2**. Drude fitting parameters for (a) CeCd$_3$P$_3$ and (b) LaCd$_3$P$_3$ as functions of temperature. (c) and (d) Comparison between measured electrical DC resistivity and that obtained from optical measurements.

To obtain more quantitative results for the metallic properties of both samples, we analyzed the optical conductivity in the low-energy region using the Drude model. In this model, the real part of the optical conductivity ($\sigma_1(\omega)$) is determined by the plasma frequency ($\Omega_{\text{Dp}}$) and the impurity scattering rate ($1/\tau$) as $\sigma_1(\omega) = \frac{\Omega_{\text{Dp}}^2}{4\pi}\frac{1/\tau}{\omega^2+(1/\tau)^2}$. The plasma frequency square is proportional to the charge carrier density ($n$), i.e., $\Omega_{\text{Dp}}^2 = \frac{4\pi n e^2}{m_b}$, where $e$ is the unit charge and $m_b$ is the band mass. Because of the dominant absorption by the IR active phonon modes between 200 and 350 cm$^{-1}$, the fitting region was restricted below 200 cm$^{-1}$, and the Drude model fits are displayed with dashed lines in Figs. 1 (c) and (d).

Figures 2 (a) and (b) show the fitting parameters of the Drude model for CeCd$_3$P$_3$ and LaCd$_3$P$_3$, respectively. The plasma frequency remained relatively constant with temperature, while the scattering rate increased gradually with temperature. The scattering rates of both samples were directly proportional to $T^2$, indicating adherence to Fermi liquid behavior. Additionally, the plasma frequencies for both samples were comparable. However, the absolute values were significantly lower than those observed in standard metals, suggesting very low charge carrier densities in both compounds. The estimated charge carrier densities of CeCd$_3$P$_3$ and LaCd$_3$P$_3$, derived from the plasma frequencies, were $2.5 \times 10^{19}$ and $2.2 \times 10^{19}$ cm$^{-3}$, respectively, consistent with values obtained from Hall resistivity measurements[32]. Note that one of the standard metals, such as copper, has a charge carrier density of $8.47 \times 10^{22}$ cm$^{-3}$. The estimation assumed the band mass to be equal to the bare electron mass. The DC resistivity values ($\rho$) are obtained from the fitting parameters of the Drude model, i.e., $\rho \equiv \frac{1}{\sigma_0} = \frac{4\pi(1/\tau)}{\Omega_{Dp}^2}$, where $\sigma_0$ is the DC conductivity. The resulting DC resistivity data are shown in Figs. 2 (c) and (d). The measured electrical DC resistivity decreased with decreasing temperature, and the optically derived DC resistivity exhibited a temperature-dependent trend similar to the measured electrical DC resistivity. However, there is a slight discrepancy between the two DC resistivities. Finding the origin of the slight discrepancy between electrical and optical resistivity is nontrivial. In the contact electrical measurement, the accurate dimensions of the specimen are necessary for obtaining DC resistivity, while, in the non-contact optical measurement, the accurate reflectance and the appropriate extrapolation to zero frequency for the Kramers-Kronig analysis are important. Therefore, the two different approaches for getting DC resistivity may naturally cause the slight discrepancy between the resulting resistivities.

Although the measured reflectance spectra of the Ce- and La-based samples, shown in Figs. 1(a) and (b), are quite similar to each other, close inspection reveals some differences. CeCd$_3$P$_3$ exhibits significant curve crossings at approximately 650 and 3000 cm$^{-1}$, whereas LaCd$_3$P$_3$ does not display these features. The insets of Figs. 1(c) and (d) present the conductivity spectra of CeCd$_3$P$_3$ and LaCd$_3$P$_3$, respectively, post subtraction of the Drude-model fit. These conductivity spectra demonstrate distinct frequency dependencies for the two samples. The only difference between the two compounds is the 4*f* electron, i.e., one has 4*f* electrons but the other does not. Therefore, the variation is likely associated with the 4*f* electrons, presumably related to the crystalline electric field effects, necessitating further investigation.

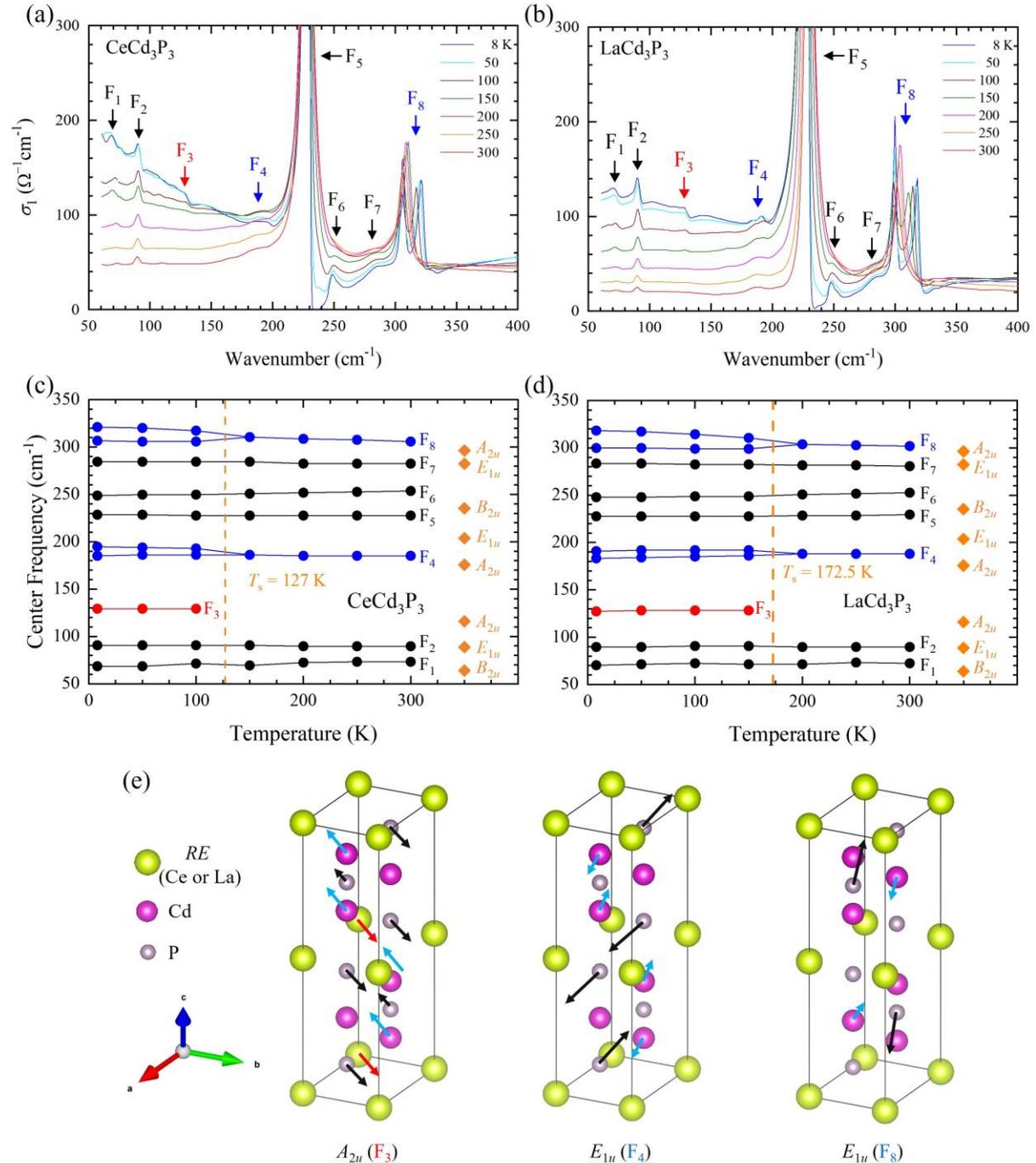

**Figure 3**. Optical conductivity spectra of (a) $CeCd_3P_3$ and (b) $LaCd_3P_3$, displaying all identified phonon modes ($F_1 - F_8$). Temperature-dependent peak positions of IR-active phonon modes of (c) $CeCd_3P_3$ and (d) $LaCd_3P_3$, along with the corresponding phonon frequencis obtained from the first-priciples calculations (Table S1). (e) Calculated vibration modes of $F_3$, $F_4$, and $F_8$ phonons, showing dominant atomic displacements.

Unlike typical metals, the extremely low carrier density of these materials enabled us to clearly observe the eight IR-active phonon modes. The phonon modes below 400 cm$^{-1}$ exhibit distinct temperature-dependent behaviors, as shown in Figs. 3 (a) and (b). The vertical arrows in the figures indicate the peak frequencies of the observed phonon modes, and the temperature dependencies of their peak positions are shown in Figs. 3 (c) and (d). In a prior study, thermodynamic and DC transport measurements revealed phase transitions near $T_s$ = 127 and 172 K for CeCd$_3$P$_3$ and LaCd$_3$P$_3$, respectively. Below $T_s$, significant modifications in the phonon spectra were observed. The phonon at 310 cm$^{-1}$ (F$_8$) slightly hardened as the temperature decreased to $T_s$ and then bifurcated below $T_s$. The phonon peak near 200 cm$^{-1}$ (F$_4$) remained relatively stable down to $T_s$ and bifurcated below $T_s$. Phonon peaks at approximately 70, 90, 230, 250, and 280 cm$^{-1}$ (F$_1$, F$_2$, F$_5$, F$_6$, and F$_7$) were present at all temperatures, while the phonon peak near 125 cm$^{-1}$ (F$_3$) existed only below $T_s$. Below $T_s$, the splitting of the peaks (F$_8$ and F$_4$) and the emergence of a new peak (F$_3$) strongly suggest a structural phase transition occurring at $T_s$ due to temperature-induced structural instability.

To interpret the observed phonon anomalies, we conducted first-principles phonon calculations at the Γ point; for computational details and a complete list of calculated phonon modes, see the Supplementary Material. The calculated IR-active phonon frequencies are plotted with the experimental data in Figs. 3 (c) and (d), and the atomic vibration modes of F3, F4, and F8 phonons, which exhibit anomalies, are illustrated in Fig. 3 (e). The illustrations highlight only the dominant atomic displacements for clarity. The observed phonon anomalies are indicative of symmetry breaking. Based on group-subgroup symmetry analysis, a possible structural transition is from the centrosymmetric *P*6$_{3/mmc}$ phase to its non-centrosymmetric maximal subgroup *P*6$_{3mc}$. This transition lifts the degeneracy of $E_{1u}$ modes and activates previously silent modes due to the loss of inversion and mirror symmetry. Our DFT results support this interpretation: the F$_4$ phonon is composed of both a *z*-polarized $A_{2u}$ mode and an in-plane $E_{1u}$ component, but only the latter is shown in Fig. 3 (e). The F$_8$ phonon originates from a degenerated $E_{1u}$ mode that splits under symmetry lowering, and the F$_3$ mode is IR-inactive in *P*6$_{3/mmc}$ but becomes active in a lower-symmetry *P*6$_{3mc}$ phase. This structural phase transition is distinct from the antiferromagnetic (AFM) phase transition of CeCd$_3$P$_3$, as the AFM transition was observed below 0.41 K. Further insights can be obtained from Raman and X-ray diffraction measurements.

**Table 1**. Lattice parameters and energy gaps of CeCd$_3$P$_3$ and LaCd$_3$P$_3$, as determined from both theoretical calculations and experiments (Refs. 32 and 39). The energy gap values ($E_g$) are obtained from Ref. 31.

|  | CeCd$_3$P$_3$ | | LaCd$_3$P$_3$ | |
| --- | --- | --- | --- | --- |
|  | Exp. | Cal. | Exp. | Cal. |
| $a$ (Å) | 4.2767 | 4.3244 | 4.2925 | 4.3365 |
| $c$ (Å) | 20.9665 | 21.2311 | 21.0763 | 21.2348 |
| Wyckoff $z_{Cd1}$ | 0.12724 | 0.12666 | 0.12724 | 0.12670 |
| Wyckoff $z_{P1}$ | 0.57775 | 0.57838 | 0.57775 | 0.57819 |
| $E_g$ (eV) | 0.75 | 0.51 | 0.73 | 0.70 |

First-principles calculations were conducted to analyze the electronic ground states of the two compounds. Table 1 presents the lattice constants and energy gaps of CeCd$_3$P$_3$ and LaCd$_3$P$_3$ derived from both experiments and theoretical calculations[31,32,39]. The theoretical values for the $a$ and $c$ lattice constants of CeCd$_3$P$_3$ and LaCd$_3$P$_3$ show a close agreement (approximately 1% difference) with the experimental values. Our calculations indicate that CeCd$_3$P$_3$ and LaCd$_3$P$_3$ are semiconductor materials with small energy gaps of approximately 0.51 and 0.70 eV, respectively. In contrast, optical spectroscopy and electrical resistivity measurements of $R$Cd$_3$P$_3$ single crystals, grown using the flux method, reveal metallic behavior.

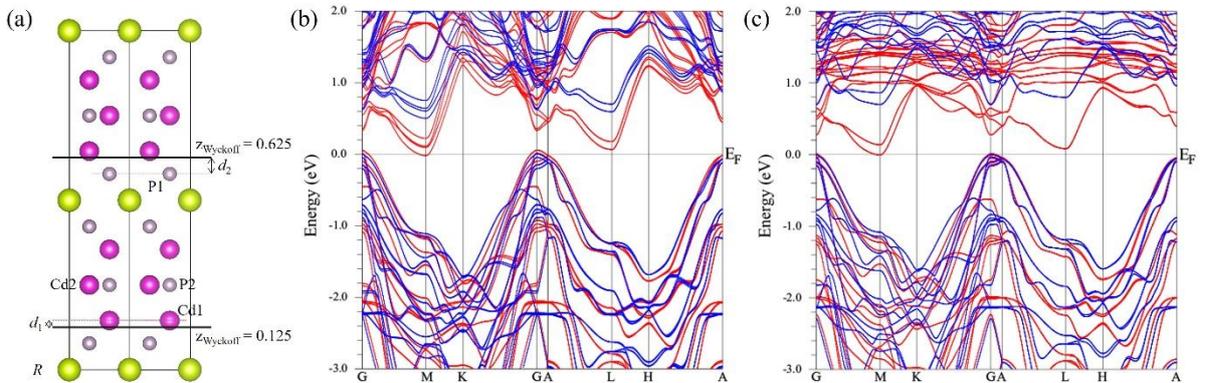

**Figure 4**. (a) Crystal structure of $R$Cd$_3$P$_3$. Electronic band structures of (b) CeCd$_3$P$_3$ and (c) LaCd$_3$P$_3$ in the metallic (red) and insulating (blue) states. Refer to the text for further details.

To investigate the discrepancy between the theoretical (insulating energy gap) and experimental (metallic behavior) results, we analyzed the band characteristics around the Fermi level of CeCd$_3$P$_3$ and LaCd$_3$P$_3$ by varying the positions of the Cd1 and P1 atoms in the crystal structure (Fig. 4(a)). In our calculations, the dominant characteristic bands of the conduction band minimum were the $R$, Cd1, and P1 atoms, while those of the valence band maximum were the Cd2 and P2 atoms (see Figs. S1 and S2 in the Supplementary Material). In the crystal structure (Fig. 4(a)), Cd2 and P2 atoms form perfect horizontal planes at the Wyckoff positions of $z_{\text{Wyckoff}} = 0.25$ and $0.75$. In contrast, Cd1 and P1 atoms are slightly displaced in opposite directions from perfect horizontal planes at the Wyckoff positions of $z_{\text{Wyckoff}} = 0.125, 0.375, 0.625$, and $0.875$. Additionally, the $R$ atoms formed perfect horizontal planes at $z_{\text{Wyckoff}} = 0.0$, $0.5$, and $1.0$. The shifts of the Cd1 and P1 atoms are significantly exaggerated in Fig. 4(a). Additionally, the shift of the light P1 atom is larger than that of the heavy Cd1 atom. The absolute values of the shifts of Cd1 and P1 atoms from the perfect plane are $d_1$ and $d_2$, respectively. We investigated the evolution of electronic band structures by shifting the Wyckoff positions of the Cd1 and P1 atoms from the perfect plane. Note that we focused on the displacements of Cd1 and P1 because they occupy the 4$f$ Wyckoff positions in the $P6_3/mmc$ space group, which allow continuous variation along the $c$-axis. In contrast, Cd2 and P2 atoms reside at high-symmetry 2$d$ and 2$c$ Wyckoff positions, respectively, with $z$-coordinates fixed by symmetry ($z = 0.75$ and $0.25$), lying on mirror planes. Therefore, displacements of Cd2 and P2 are not allowed within the parent symmetry, and we restricted our analysis to Cd1 and P1. The theoretical values of $d_1$ and $d_2$ are assumed to be their maximum values, i.e., $\Delta_1$ and $\Delta_2$, respectively. When both $d_1$ and $d_2$ are zero, the two atoms are exactly on the same plane, implying that the Cd1 and P1 atoms form perfect planes at $z_{\text{Wyckoff}} = 0.125, 0.375, 0.625$, and $0.875$. We introduced an adjustable parameter $\alpha$, which can be any number between $0.0$ and $1.0$. Values of $d_1$ and $d_2$ between $0.0$ and their maxima can be expressed using the parameter $\alpha$ as $d_1 = \alpha \Delta_1$ and $d_2 = \alpha \Delta_2$, respectively. When $\alpha = 1.0$, the positions of the Cd1 and P1 atoms are the same as the theoretical values in Table 1. When $\alpha = 0.0$, the Cd1 and P1 atoms form a perfect plane. The positions of the Cd1 and P1 atoms can be gradually shifted from the perfect plane to the theoretical positions by adjusting the parameter from $0.0$ to $1.0$ in increments of $0.05$. Table 2 shows the Wyckoff positions of Cd1 and P1 as functions of $\alpha$ along the $z$-axis. This type of coordinated out-of-plane displacement of Cd1 and P1 atoms can be viewed as a static breathing-like distortion, in which atoms move symmetrically toward or away from the mirror planes. Although we did not perform explicit phonon calculations, such a phonon mode is a plausible candidate for describing the observed symmetry-lowering behavior. Figures 4(b) and (c) show the metal–insulator transitions of CeCd$_3$P$_3$ and LaCd$_3$P$_3$, which occur between two adjacent $\alpha$ values. The metal-insulator transitions take place between 0.80 and 0.85 for CeCd$_3$P$_3$ and between 0.75 and 0.80 for LaCd$_3$P$_3$. When the metal-insulator transition occurs, the differences between the Wyckoff positions of the Cd1 and P1 atoms of CeCd$_3$P$_3$ (LaCd$_3$P$_3$) are about 1.766 (1.806) × 10$^{-3}$ and 4.949 (4.971) × 10$^{-2}$ Å in the unit cell, respectively. It is challenging to detect these small differences experimentally. The minute changes in position are within the experimental error bar. Minute shifts from the actual positions of the Cd1 and P1 atoms in the crystal structure may explain the metallic behavior observed in the optical and electrical experiments. Our results suggest a generic in-plane structural instability for this family of materials, which warrants further structural investigation. In addition, our results motivated further experiments to determine whether the in-plane lattice instability was related to the unidentified structural transition at $T_s$ in metallic $R$Cd$_3$P$_3$ and $R$Cd$_3$As$_3$ crystals.

**Table 2**. Metal–insulator transition. Energy gaps and Wyckoff positions of the Cd1 and P1 in the unit cells of CeCd$_3$P$_3$ and LaCd$_3$P$_3$.

| CeCd$_3$P$_3$ | | | | LaCd$_3$P$_3$ | | | |
|---|---|---|---|---|---|---|---|
| $\alpha$ | $E_g$ (eV) | $z_{Cd1}$ | $z_{P1}$ | $\alpha$ | $E_g$ (eV) | $z_{Cd1}$ | $z_{P1}$ |
| 0 | 0 (metal) | 0.125 | 0.625 | 0 | 0 (metal) | 0.125 | 0.625 |
| … | 0 (metal) | … | … | … | 0 (metal) | … | … |
| 0.8 | 0 (metal) | 0.12633102 | 0.58770222 | 0.75 | 0 (metal) | 0.12627575 | 0.58988875 |
| 0.85 | 0.075 (insul.) | 0.12641420 | 0.58537111 | 0.8 | 0.116 (insul.) | 0.12636080 | 0.58754800 |
| … | … (insul.) | … | … | … | … (insul.) | … | … |
| 1 | 0.51 (insul.) | 0.12666 | 0.57838 | 1 | 0.7 (insul.) | 0.12670 | 0.57819 |

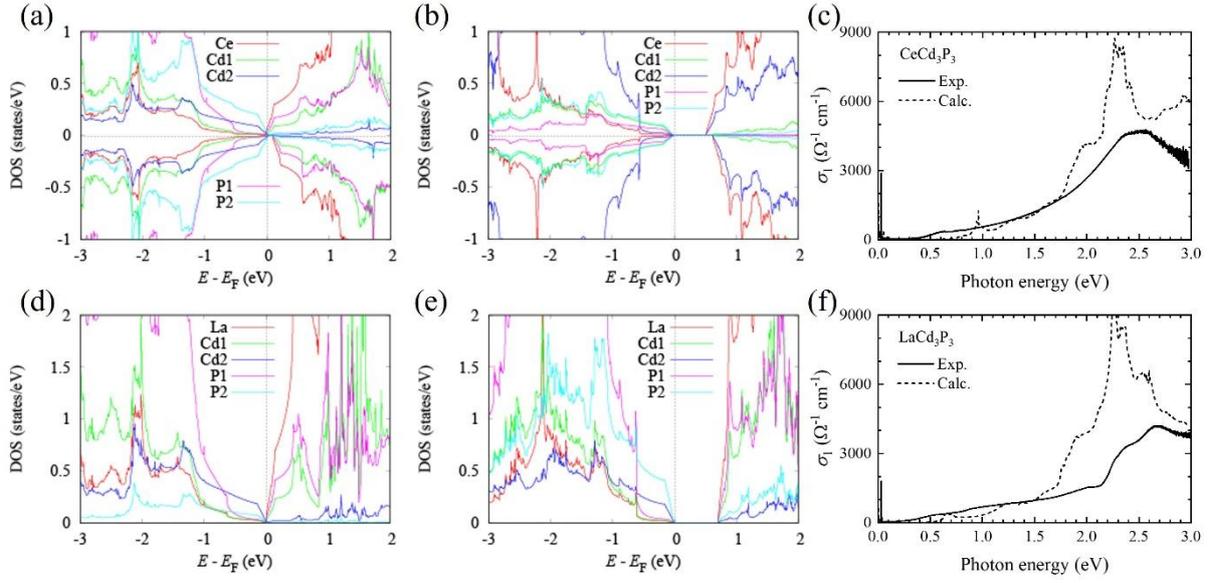

**Figure 5.** Density of states (DOS) for (a, d) metallic and (b, e) insulating phases of $CeCd_3P_3$ and $LaCd_3P_3$ compounds. Experimental and theoretical optical conductivity spectra of (c) $CeCd_3P_3$ and (f) $LaCd_3P_3$ compounds.

The density of states (DOS) of $CeCd_3P_3$ and $LaCd_3P_3$ was computed to understand the low charge carrier density observed in optical experiments. Figures 5 (a) and (b) show the DOS of the metallic and insulating phases of $CeCd_3P_3$, while Figs. 5 (d) and (e) present the DOS of the metallic and insulating phases of $LaCd_3P_3$, respectively (see Figs. 4(b) and (c)). In the metallic phase, because the DOS at the Fermi level is significantly small, a low carrier density is expected. Furthermore, the optical conductivity spectra of $CeCd_3P_3$ and $LaCd_3P_3$ compounds were determined from theoretical results and compared with experimentally measured optical conductivity spectra, depicted in Figs. 5 (c) and (f). The experimental optical conductivity was acquired at the lowest measured temperature of 8 K, while the calculated optical conductivity spectra were derived for the metallic phase. The experimental and theoretical conductivity spectra exhibit similarity, providing further evidence supporting the unique electronic characteristics of these materials and enhancing our understanding of their optical properties.

Our findings indicate that measuring electrical resistivity under uniaxial strain can be a method to systematically control the insulator-metal transition in two-dimensional $RCd_3X_3$ ($X$ = P and As) materials. Discrepancies between previous studies[31,32], show that flux-grown single crystals of $RCd_3P_3$ exhibit metallic behavior in optical and electrical results. On the other hand, polycrystalline samples exhibit semiconducting behavior due to slight differences in atomic positions of Cd1 and P1 atoms in the unit cell. The same rationale applies to $R$Cd$_3$As$_3$. It has been demonstrated that the Kondo effect can be induced in a *p*-type semiconductor, CeZn$_3$P$_3$, by illumination with visible light[29,30]. Compared to other non-artificial Kondo effects, the photoinduced Kondo effect occurs at higher temperatures and extends the scope of the device, providing a range of potential operations for magneto-optical devices, as well as quantum information/computation devices[40-47]. The observation of conventional metallic behavior in terms of electrical resistivity and optical conductivity indicates that the 4*f* electrons of CeCd$_3$P$_3$ are negligibly hybridized with the conduction electrons. Furthermore, CeCd$_3$P$_3$ can be used as a platform for photoinduced Kondo effect materials because its ground state is a metallic phase with small charge carriers. If sufficient charge carriers are supplied by light illumination, this material is expected to exhibit the Kondo effect. Thus, CeCd$_3$P$_3$ provides an opportunity to study the complex interactions between magnetic frustration and the Ruderman–Kittel–Kasuya–Yosida interaction in triangular Ce lattices with low carrier densities, as well as the photoinduced Kondo effect.

**CONCLUSION**

We examined the electronic structures and optical properties of $R$Cd$_3$P$_3$ ($R$ = Ce and La) compounds. Our first-principles calculations indicate that CeCd$_3$P$_3$ and LaCd$_3$P$_3$ exhibit semiconducting ground states with narrow energy gaps. However, optical spectroscopy revealed that both compounds behave as metals with significantly low charge carrier densities. Moreover, a significant modification in the phonon spectra in the IR region suggests a structural phase transition in these materials. Notably, a minor adjustment of the Cd1 and P1 Wyckoff positions induces a metal–insulator transition, elucidating the metallic behavior observed experimentally in both compounds.

**METHODS**

**Sample preparation and optical measurements**

Single crystals of $R$Cd$_3$P$_3$ ($R$ = La or Ce) were synthesized using high-temperature ternary melt[48]. The detailed physical properties of $R$Cd$_3$P$_3$ ($R$ = La and Ce), including magnetization, electrical resistivity, Hall coefficient, and specific heat, are available in the literature[32]. The sample dimensions are $3 \times 3$ mm$^2$. Reflectance spectra in a wide spectral range (60–25000 cm$^{-1}$) were measured using a commercial spectrometer (Vertex 80v, Bruker) at temperatures ranging from 300 K down to 8 K. The optical conductivity was determined from the measured reflectance using Kramers–Kronig analysis[37,38]. To conduct the Kramers–Kronig integration, the reflectance spectrum measured within a finite spectral range needs to be extrapolated to zero and infinity. For extrapolation from the lowest data point to zero, the Hagen–Rubens relation, i.e., $1 - R(\omega) \propto \sqrt{\omega}$. For extrapolation from the highest data point to infinity, from the highest data point (25000 cm$^{-1}$) to $10^6$ cm$^{-1}$, we used $R(\omega) \propto \omega^{-2}$, above $10^6$ cm$^{-1}$, we assumed the free electron behavior, i.e., $R(\omega) \propto \omega^{-4}$.

**Electronic structure calculations**

The Quantum ESPRESSO (QE) package[49] was used for performing structural optimization calculations. The lattice constants were fully relaxed using experimental values. The projector-augmented wave method, combined with the Perdew–Burke–Ernzerhof (PBE)[50] exchange-correlation functional, was employed without incorporating SOC. The cutoff energies for wave functions and charge density were set at 40 Ry and 640 Ry for CeCd$_3$P$_3$ and 75 Ry and 640 Ry for LaCd$_3$P$_3$, respectively. Convergence thresholds were set to $1.0 \times 10^{-8}$ Ry for total energy and $1.0 \times 10^{-7}$ Ry/Bohr for forces, with a pressure convergence threshold of 0.05 kbar. Brillouin-zone sampling was performed using a 10×10×2 Monkhorst-Pack k-point mesh. The experimental lattice constants were referenced from Ref. 32.

WIEN2k package[51] implemented with the full-potential linearized-augmented-plane-wave method was used for electronic structure calculations. The generalized gradient approximation of the PBE was selected as the exchange-correlation functional. The parameter $RK_{max}$ was set to 8, with muffin-tin radii of 2.5 a.u. for Ce, La, and Cd atoms, and 2.11 a.u. for P atoms. For the self-consistent-field cycle (SCF), DOS, and optical conductivity, we generated 3000 and 50000 k-points in the whole Brillouin zone (BZ), corresponding to 195 and 2310 k-points in an irreducible wedge of BZ, respectively. The convergence criterion for total energy during SCF cycles was set to $1 \times 10^{-7}$ Ry. The calculations included orbital-dependent potentials using the LDA+$U$ method[52] and SOC. Incorporating the LDA+$U$ method for the Ce 4$f$ orbital involved setting the on-site Coulomb interaction ($U$) to approximately 6 eV and the exchange parameter $J$ to 0, resulting in the effective potential $U_{eff} = U - J$. Note that we employed the QE software package for the structural optimization, while the electronic structure calculation was completed in the WIEN2k software package. This hybrid approach allowed us to combine the computational efficiency of QE for structural relaxation with the high accuracy and rigor of WIEN2k for treating correlated electron systems, ultimately providing a more reliable and physically meaningful understanding of the materials' electronic properties.

The optical conductivity spectrum was determined using $\sigma(\omega) = \frac{\pi e^2}{3m^2\omega}\sum_{f,i}\int_{BZ}d^3k\frac{2}{(2\pi)^3}|\boldsymbol{P}_{fi}|\delta[E_f(\boldsymbol{k})-E_i(\boldsymbol{k})-\hbar\omega]$, where $m$ and $e$ are the electronic mass and the unit charge, respectively, $\hbar\omega$ is the photon energy, $E_i(\boldsymbol{k})$ and $E_f(\boldsymbol{k})$ are the energies of the initial and final states, respectively, $\boldsymbol{k}$ is the wave vector inside the BZ where the transition from $E_i(\boldsymbol{k})$ to $E_f(\boldsymbol{k})$ occurs, and the optical transition-matrix element is $\boldsymbol{P}_{fi}=\frac{\hbar}{i}\langle f|\boldsymbol{\nabla}|i\rangle$. The initial (final) state $|i\rangle$ ($|f\rangle$), is occupied (unoccupied). Here, it is assumed that only direct transitions occur.

## ACKNOWLEDGMENTS

This study received funding from the Basic Science Research Program of the National Research Foundation of Korea (NRF), sponsored by the Ministry of Education. J.H. and Y.S.S. acknowledge financial support from the National Research Foundation of Korea (NRFK Grant Nos. 2021R1A2C101109811, 2022R1I1A1A01068619, and RS-2024-00460248). E.M. acknowledges support from the Natural Sciences and Engineering Research Council of Canada, Canada Research Chairs, and the Canada Foundation for Innovation program.

## AUTHOR CONTRIBUTIONS

J.J. and Y.S.S. are co-first authors. The single crystals were grown by J. L. and E.M. J.J. performed the DFT calculations. Optical measurements and analysis were performed by Y. S. S. and J. H. All authors discussed the results and interpretations. J.J., Y.S.S., E.M., and J.H. wrote the manuscript with input from all authors.

## COMPETING INTERESTS

The authors declare no competing interests.

## DATA AVAILABILITY

All data supporting the findings of this study are available from the corresponding authors upon request.

**Table 2**. Lattice parameters and energy gaps of $CeCd_3P_3$ and $LaCd_3P_3$, as determined from both theoretical calculations and experiments (Refs. 32 and 39). The energy gap values ($E_g$) are obtained from Ref. 31.

|  | $CeCd_3P_3$ | | $LaCd_3P_3$ | |
| --- | --- | --- | --- | --- |
|  | Exp. | Cal. | Exp. | Cal. |
| $a$ (Å) | 4.2767 | 4.3244 | 4.2925 | 4.3365 |
| $c$ (Å) | 20.9665 | 21.2311 | 21.0763 | 21.2348 |
| Wyckoff $z_{Cd1}$ | 0.12724 | 0.12666 | 0.12724 | 0.12670 |
| Wyckoff $z_{P1}$ | 0.57775 | 0.57838 | 0.57775 | 0.57819 |
| $E_g$ (eV) | 0.75 | 0.51 | 0.73 | 0.70 |

**Table 2**. Metal–insulator transition. Energy gaps and Wyckoff positions of the Cd1 and P1 in the unit cells of $CeCd_3P_3$ and $LaCd_3P_3$.

| $CeCd_3P_3$ | | | | $LaCd_3P_3$ | | | |
| --- | --- | --- | --- | --- | --- | --- | --- |
| $\alpha$ | $E_g$ (eV) | $z_{Cd1}$ | $z_{P1}$ | $\alpha$ | $E_g$ (eV) | $z_{Cd1}$ | $z_{P1}$ |
| 0 | 0 (metal) | 0.125 | 0.625 | 0 | 0 (metal) | 0.125 | 0.625 |
| … | 0 (metal) | … | … | … | 0 (metal) | … | … |
| 0.8 | 0 (metal) | 0.12633102 | 0.58770222 | 0.75 | 0 (metal) | 0.12627575 | 0.58988875 |
| 0.85 | 0.075 (insul.) | 0.12641420 | 0.58537111 | 0.8 | 0.116 (insul.) | 0.12636080 | 0.58754800 |
| … | … (insul.) | … | … | … | … (insul.) | … | … |
| 1 | 0.51 (insul.) | 0.12666 | 0.57838 | 1 | 0.7 (insul.) | 0.12670 | 0.57819 |

# Supplementary Material

**1. Electronic band structure calculations**

(a)

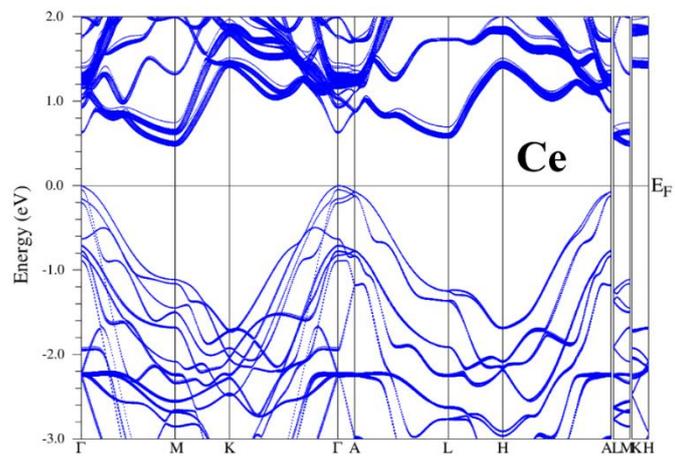

(b)

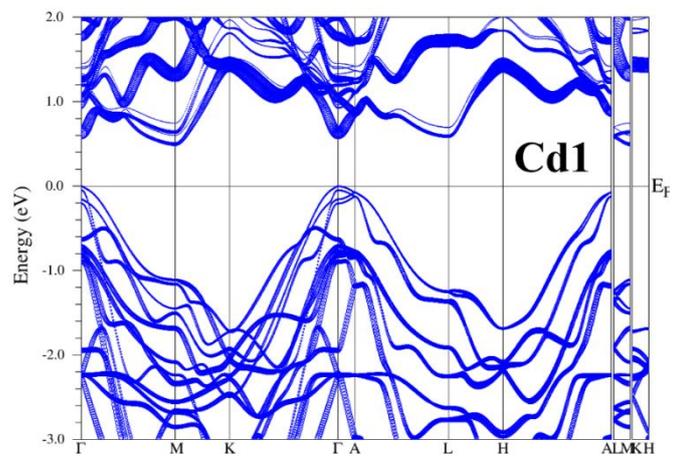

(c)

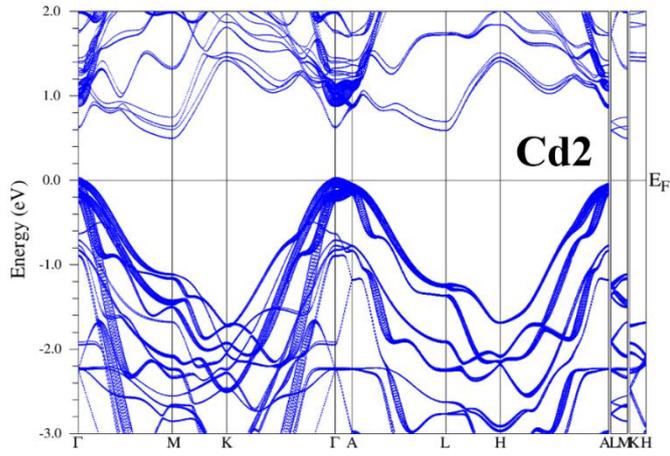

(d)

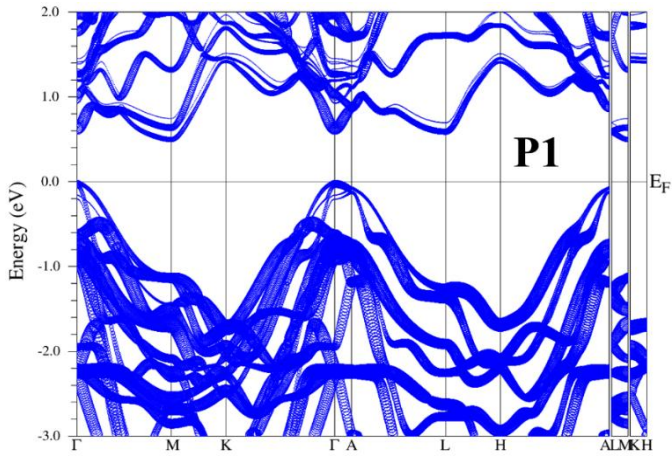

(e)

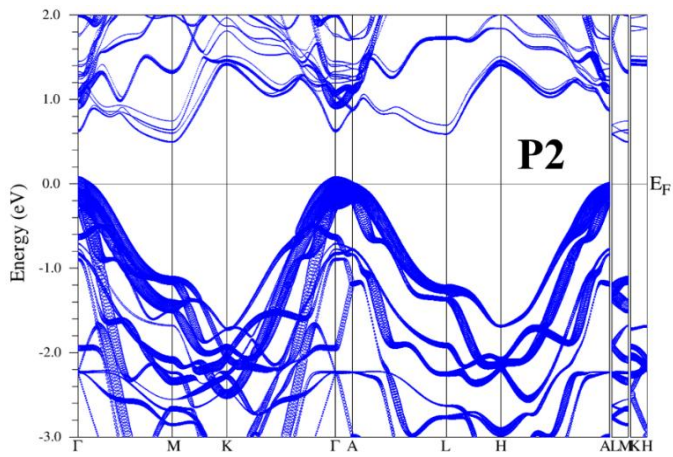

**Figure S1.** Characteristic band structures for Ce (a), Cd1 (b), Cd2 (c), P1 (d), and P2 (e) atoms of CeCd$_3$P$_3$ compound in insulating state.

(a)

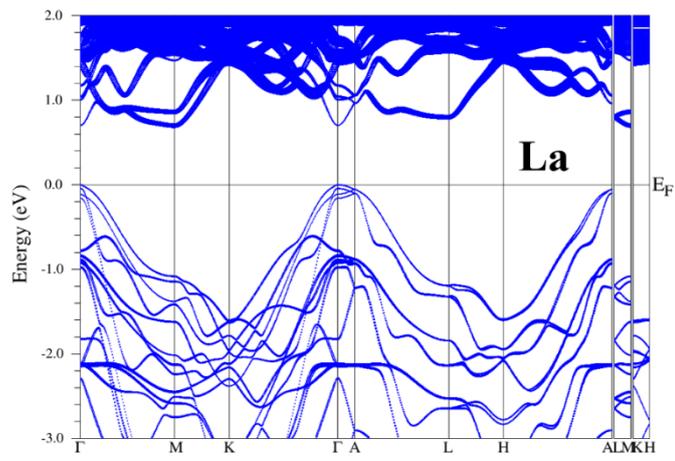

(b)

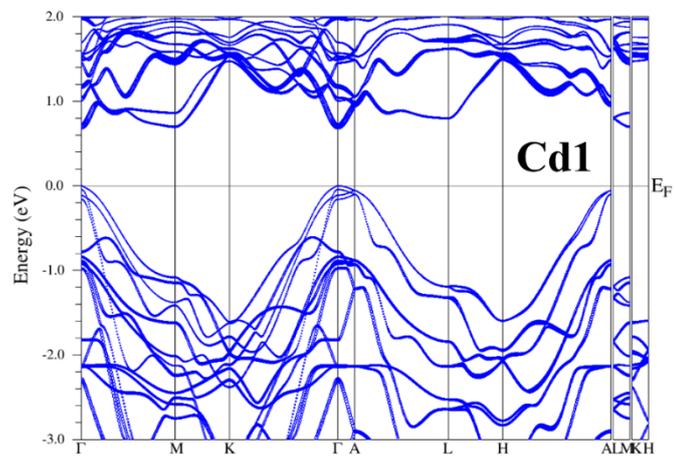

(c)

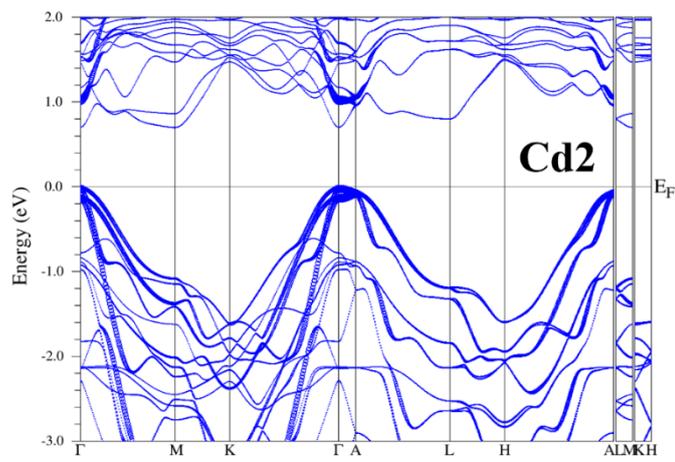

(d)

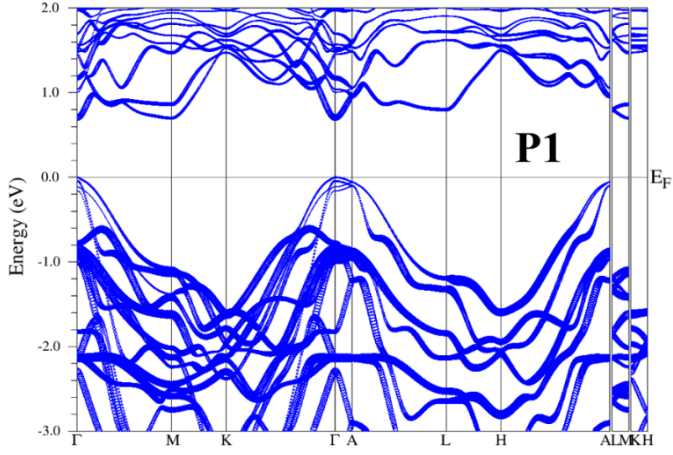

(e)

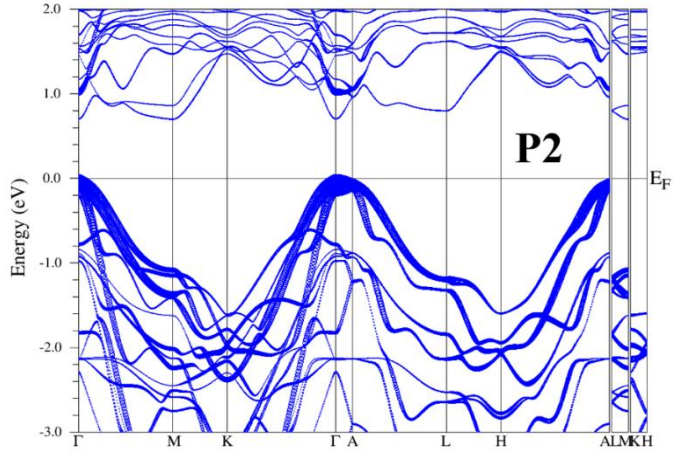

**Figure S2.** Characteristic band structures for La (a), Cd1 (b), Cd2 (c), P1 (d), and P2 (e) atoms of LaCd$_3$P$_3$ compound in insulating state.

Figures S1 and S2 represent the characteristic band structures of $R$Cd$_3$P$_3$ compounds in an insulating state. According to our calculations, the dominant characteristic bands of the conduction band minimum are $R$, Cd1, and P1 atoms, and those of the valence band maximum are Cd2 and P2 atoms.

## 2. Phonon analysis based on first-principles calculations

### 2.1 Computational details

Phonon dispersion relations for the CeCd$_3$P$_3$ compound were computed using density functional perturbation theory (DFPT) as implemented in the Quantum ESPRESSO package. The electronic ground state required for the DFPT calculations was obtained using a 9×9×6

Monkhorst–Pack *k*-point grid, providing accurate sampling of the Brillouin zone. A kinetic energy cutoff of 40 Ry was used for the plane-wave expansion of the electronic wavefunctions, and a higher cutoff of 640 Ry was applied for the charge density and potential, as required by the PAW-type pseudopotentials to ensure reliable numerical accuracy.

Based on the converged charge density, the dynamical matrices were calculated on a uniform 3×3×3 *q*-point mesh. A tight self-consistency threshold of $1.0\times10^{-14}$ was employed to ensure the accurate evaluation of the perturbative response. To enhance numerical stability during the phonon calculations, a moderate charge-density mixing scheme was used. The acoustic sum rule was enforced by symmetrizing the dynamical matrices, guaranteeing that the acoustic modes correctly vanish at the *Γ*-point. The resulting phonon frequencies were interpolated along high-symmetry paths in the Brillouin zone to obtain the full phonon dispersion relations.

## 2.2 List of calculated *Γ*-point phonon modes

To compare with the experiment, we selectively list the *Γ*-point phonon modes that correspond to the experimentally observed IR-active peaks ($F_1$–$F_8$). The selected modes are summarized in Table S1 along with their calculated frequencies, symmetry labels, IR activity, dominant atomic contributions, and displacement directions.

The $F_1$ (~70 cm$^{-1}$) and $F_5$-$F_6$ (~230-250 cm$^{-1}$) phonons are assigned to $B_{2u}$-type modes, which are nominally IR-inactive but may become weakly allowed due to symmetry lowering. These modes involve *z*-polarized displacements of atoms such as Ce and P2 ($F_1$) and Cd1 and P1 ($F_5$-$F_6$), suggesting that inversion symmetry loss may activate these otherwise silent modes.

**Table S1.** Assignment of *Γ*-point phonon modes to experimental IR-active peaks ($F_1$-$F_8$)

| Mode # | Exp. Freq. (cm$^{-1}$) | Calc. Freq. (cm$^{-1}$) | Symmetry | Activity | Dominant Atoms | Dominant Direction |
|---|---|---|---|---|---|---|
| $F_1$ | ~ 70 | 64.0 | $B_{2u}$ | - | Ce, P2 | *z* |
| $F_2$ | ~ 90 | 88.7 | $E_{1u}$ | IR | Cd2, P1 | *x, y* |
| $F_3$ | ~ 125 (below $T_s$) | 116.0 | $A_{2u}$ | IR | Cd2, P1 | *x, y* |
| $F_4$ (low) | ~ 193 | 175.6 | $A_{2u}$ | IR | Ce, P2 | *z* |
| $F_4$ (high) | ~ 228 | 203.8 | $E_{1u}$ | IR | Cd1, P2 | *x, y* |
| $F_5$ & $F_6$ | ~ 230-250 | 235.4 | $B_{2u}$ | - | Cd1, P1 | *z* |
| $F_7$ | ~284 | 282.4 | $A_{2u}$ | IR | Ce | *z* |

| F$_8$ | ~310 | 296.2 (split) | $E_{1u}$ | IR | Cd2, P1 | x, y |

## 2.3 Phonon dispersion curve

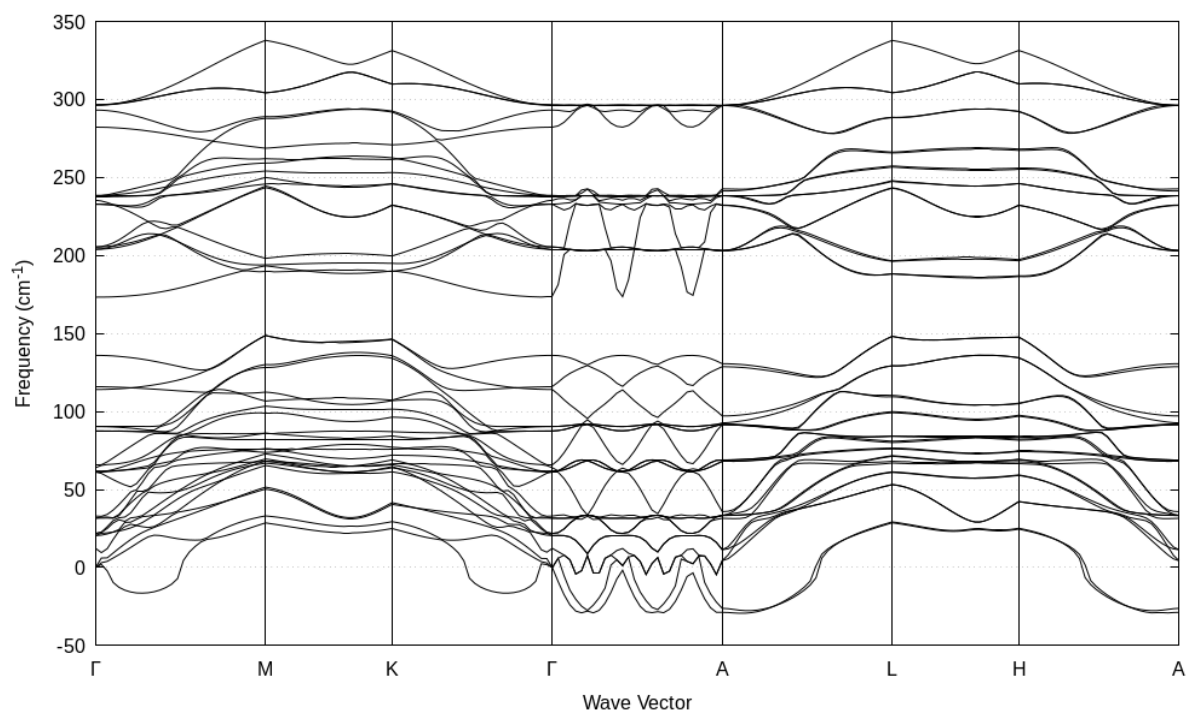

**Figure S3.** Calculated phonon dispersion curve of CeCd$_3$P$_3$ in the high-temperature P6$_{3/mmc}$ phase.

The phonon dispersion was calculated along high-symmetry directions of the hexagonal Brillouin zone, assuming the path begins at the $\Gamma$ point. As shown in Fig. S3, all phonon branches remain real throughout the zone except for a slight imaginary frequency near $\Gamma$, suggesting a tendency toward structural instability at ambient conditions.